\documentclass[pra,aps,twocolumn,superscriptaddress,floatfix,noshowpacs,nofootinbib,10pt]{revtex4-2}

\usepackage[utf8]{inputenc}
\usepackage{microtype}
\usepackage{amsmath,amssymb}
\usepackage{graphicx}
\usepackage[colorlinks=true]{hyperref}
\usepackage{color}
\usepackage{graphicx}
\usepackage[normalem]{ulem}

\newcommand{\bs}[1]{{\boldsymbol{#1}}}
\newcommand{\bk}{\bs{k}}

\newcommand{\br}{\bs{r}}
\newcommand{\bv}{\bs{v}}
\newcommand{\eps}{\bs{\epsilon}}

\begin{document}

\title{Weak localization of light in hot atomic vapors}

\author{N. Cherroret}
\email{cherroret@lkb.upmc.fr}
\affiliation{Laboratoire Kastler Brossel, Sorbonne Universit\'{e}, CNRS, ENS-PSL Research University, 
Coll\`{e}ge de France; 4 Place Jussieu, 75005 Paris, France}

\author{M. Hemmerling}
\affiliation{Instituto de F\'{i}sica de S\~ao Carlos, Universidade de S\~ao Paulo, 13560-970 S\~ao Carlos, S\~ao Paulo, Brazil}
\affiliation{Universit\'e C\^ote d'Azur, CNRS, Institut de Physique de Nice, Valbonne F-06560, France}

\author{G. Labeyrie}
\affiliation{Universit\'e C\^ote d'Azur, CNRS, Institut de Physique de Nice, Valbonne F-06560, France}

\author{D. Delande}
\affiliation{Laboratoire Kastler Brossel,
Sorbonne Universit\'{e}, CNRS, ENS-PSL Research University, 
Coll\`{e}ge de France; 4 Place Jussieu, 75005 Paris, France}

\author{J.T.M. Walraven}
\affiliation{Van der Waals-Zeeman Institute, Institute of Physics, University of Amsterdam, Science Park 904, 1098 XH Amsterdam, The Netherlands}

\author{R. Kaiser}
\affiliation{Universit\'e C\^ote d'Azur, CNRS, Institut de Physique de Nice, Valbonne F-06560, France}

\begin{abstract}
We theoretically explore the possibility to detect weak localization of light in a hot atomic vapor, where one usually expects the fast thermal motion of the atoms to destroy any interference in multiple scattering. 
To this end, we compute the coherent backscattering peak, assuming high temperature and taking into account the quantum level structure of the atomic scatterers. 
It is found that the decoherence due to thermal motion can be partially counterbalanced by  working at large laser detuning and  using small atomic cells with an elongated geometry. Under these conditions, our estimates suggest that weak localization in a hot vapor should be within reach of experimental detection.
\end{abstract}

\maketitle

\section{Introduction}
\label{sec:introduction}

Light propagating through dilute ensembles of scattering objects gives rise to interference between the fields emitted by individual scatterers. As is well known, this interference pattern is in general not a purely random superposition of fields but may exhibit non-trivial features that survive even after an ensemble average is performed.
Those include the coherent backscattering (CBS) effect \cite{Albada85, Wolf85}, a manifestation of weak localization, long-range speckle correlations \cite{Feng91, Berkovits94} or enhanced intensity fluctuations \cite{Nieuwenhuizen95} to give a few examples.  
Scattering of light is also investigated in cold atomic vapors, where CBS can be observed as well \cite{Labeyrie1999}. In these systems, special attention has been paid to dense atomic clouds, where the atoms no longer emit independently due to dipole-dipole interactions. This impacts the optical properties of atomic vapors, leading to resonance shifts or broadening \cite{Keaveney12, Javanainen14, Jennewein16}, modifies the mean free path and refractive index \cite{Cherroret16}, and affects the transmission properties of light \cite{Saint-Jalm18, Pellegrino14, Corman17, Guerin16, Ferioli21}. Dipole-dipole interactions also compete with the phenomenon of Anderson localization, to date not yet observed for light in three dimensions \cite{Skipetrov14}.

In all descriptions of light scattering in an atomic vapor, an important ingredient is the temperature. While interference or collective effects are fully at play in a cold atomic gas, this is no longer the case in a thermal cloud, where the atomic positions are not frozen. 
A central motivation for studying light scattering in clouds of moving atoms is to clarify how decoherence emerges when turning from a cold gas to a hot, ``classical'' one. Recently this problem has triggered much interest in the context of collective scattering \cite{Peyrot19, Ribeiro21}, metrology \cite{Sedlacek13}, and quantum information \cite{Whiting17}. A typical illustration of the role of decoherence is provided by the weak localization effect, which corresponds to the interference between two optical  paths involving the  same sequence of atomic scattering events but traveled in opposite directions. When the atoms are moving, the two  interfering paths display Doppler frequency shifts that manifest themselves  as an effective decoherence mechanism suppressing weak localization \cite{Golubentsev84, Snieder06, Labeyrie06}.  In hot atomic vapors, the Doppler effect 
may also lead to a broad distribution of mean free paths, triggering a phenomenon of anomalous diffusion \cite{Mercadier09}. 
To explore the impact of thermal motion  on coherent light scattering, atomic scatterers offer specific advantages compared to classical ones, such as colloids in suspension. A first one is the possibility to exploit the laser detuning to turn in a controlled fashion from a regime of resonant scattering to a far, off-resonant situation. 
Another one is the typical large value of the scatterer velocities in an atomic cloud, which makes the role of decoherence significant even at low temperature. This is in stark contrast to the velocities of classical scatterers, always much smaller due to the larger mass of the scattering objects. In light scattering from suspended colloids or teflon particles for instance, the velocities involved are such that the motion of the scatterers is usually irrelevant from the point of view of CBS decoherence, even at room temperature.

While coherent light scattering  at finite temperature has been previously explored in cold gases, where the Doppler effect remains moderate, so far hot atomic clouds have received less attention. A main reason is the naive expectation that in a hot atomic vapor, the fast thermal motion of the atoms is strong enough to destroy any interference in multiply scattered light. Recently, however, it was shown that interference involving scattering from an atom and a mirror (``mirror-assisted coherent backscattering'' \cite{Moriya16, Piovella17}) could survive thermal motion in a hot vapor \cite{Cherroret19}. To achieve this goal, the main idea was to work in a large detuning regime, where scattering occurs on a time scale that is faster than the time taken by an atom to move over an optical wavelength. 
In the present paper, we extend this idea to the mechanism  of weak localization, by theoretically exploring the possibility to detect coherent backscattering of light in a hot atomic gas. 
We find that the decoherence associated with  Doppler shifts is indeed limited at large detuning. 
Our theory further shows that the main contribution to CBS in a hot vapor stems from pairs of atoms close to each other and aligned with the optical axis, and therefore can be enhanced by using small, preferably elongated atomic cells. Under these conditions, we provide realistic theoretical estimates for the properties of the CBS peak (contrast and width) and compare them to numerical simulations, taking into account the atomic quantum level structure. These estimates suggest that CBS should be measurable in a hot vapor.

The article is organized as follows. 
In Sec.~\ref{sec_cold_atoms}, we recall the standard theoretical description of CBS in the presence of moving  atoms in a cold vapor. The case of a hot vapor in then addressed in Sec. \ref{sec_scalar_CBS} where, for the sake of pedagogy, we model the atoms by classical dipoles and adopt a scalar description of light scattering. In Sec. \ref{sec:quantum_structure}, we generalize this approach to the case of a realistic atomic transition, properly taking into account the quantum level structure and the vector character of light. In Sec. \ref{sec:geometry}, we then show how the use of smaller atomic cells, preferably of elongated shape, eventually allows one to significantly enhance the contrast of the CBS peak while maintaining a large enough weak localization signal. We finally summarize our results in Sec.~\ref{sec:conclusion}.

\section{Cold atoms}
\label{sec_cold_atoms}

Although we are primarily concerned with light scattering in a hot vapor, it is instructive to first recall how thermal motion affects coherent backscattering   in a cold atomic gas. This problem, previously investigated experimentally \cite{Labeyrie06}, indeed provides insight on the strategy to counteract the impact of decoherence on weak localization when turning to higher temperatures. 
 For the sake of clarity, here we deliberately stay at a qualitative level. 
 Further details about the description of temperature effects on coherent backscattering in cold gases can be found in \cite{Labeyrie06, Golubentsev84, Snieder06}. 

\begin{figure}[h]
\includegraphics[scale=0.75]{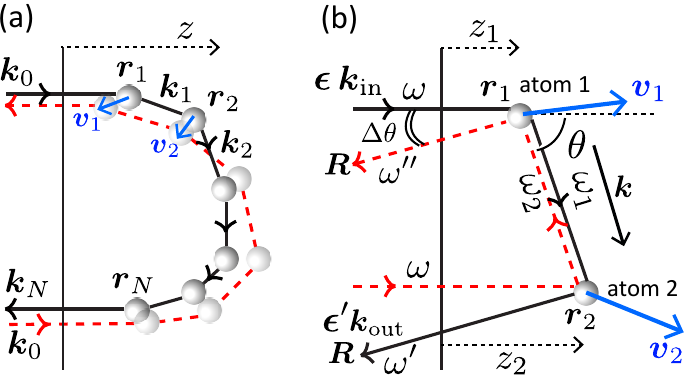}
\caption{
\label{fig:CBS_scheme}
(a) Interference between two counter-propagating scattering  paths responsible for coherent backscattering of light in a cold atomic gas at nonzero temperature (arrows on paths refer to the direction of propagation). In a temporal picture, each atom moves a little in between the scattering events on the direct and reversed paths.
(b) In a hot vapor, the main contribution to coherent backscattering involves two atoms only. In the frequency domain, thermal motion leads to a Doppler frequency shift between the paths (see main text).
}
\end{figure}

Anticipating the case of a hot atomic vapor, usually contained in a slab-shaped cell, we consider the geometry of a semi-infinite  atomic medium illuminated  at normal incidence by a plane-wave laser beam of frequency $\omega$ and wave number $k=\omega/c$.  As is well known, CBS stems from the interference between two wave amplitudes associated with multiple scattering sequences that are identical but traveled in opposite directions. This process is illustrated in Fig. \ref{fig:CBS_scheme}(a), for a given scattering sequence of $N$ atoms, located at points $\br_1,\ldots,\br_N$ in the half space $z>0$. At each scattering event, the optical wave vector is randomized, following the sequence $\bk_0,\ldots,\bk_N$. In this section, we focus on the CBS contrast at backscattering, thus choosing $\bk_N=-\bk_0$.
For this sequence of $N$ scattering events, the CBS contrast, $c(N)$, is defined as the ratio of the interference contribution in Fig. \ref{fig:CBS_scheme}(a) to the corresponding incoherent diagram where the two paths follow the same sequence in the same order.
 At zero temperature, the atomic positions are fixed, so that no dephasing occurs and the interference is fully visible, $c(N)=1$. In contrast, at finite temperature an atom located at $\br_n$ moves during the time window 
 that separates the scattering events on the direct and reversed paths. 
 Specifically, if one denotes by $\tau$ the time taken by the wave to scatter from an atom to the next one,
the direct and reversed paths reach atom $n$ at a time $n\tau$ and $(N-n+1)\tau$, respectively. 
The dephasing beween the two trajectories is then
\begin{equation}
\label{eq_dephasing_cold}
\Delta\Phi_N=\sum_{n=1}^N (\bk_n\!-\!\bk_{n-1})\cdot[\br_n(n\tau)\!-\!\br_n((N\!-\!n\!+\!1)\tau)].
\end{equation}
The CBS contrast of the sequence is given by 
\begin{equation}
c(N)=\langle\exp(i\Delta\Phi_N)\rangle,
\end{equation}
where the brackets refer to averaging over the atomic velocities and over the scattering directions $\smash{\hat{\bk}_i=\bk_i/k}$.
At the temperatures involved in experiments on cold gases, the atoms typically move ballistically, so that $\br_n(t)=\bv_n t$, with $\bv_n$ the velocity of atom $n$. We assume these velocities to follow a Gaussian distribution (cloud in thermal equilibrium),
\begin{equation}
p(\bv)=\frac{1}{(\sqrt{2\pi} \bar{v})^3}\exp\left(-\frac{\bv^2}{2\bar{v}^2}\right),
\label{velocity_distrib}
\end{equation}
where $\bar{v}$ is the one-dimensional rms speed, related to the temperature through $k_BT=m\bar{v}^2$. Performing the average over the $\bv_n$ using Eq. (\ref{velocity_distrib}), we obtain
\begin{equation}
c(N)\!=\!\left\langle\!\exp\!\bigg[\!-\!\frac{k^2\bar{v}^2\tau^2}{2}\sum_{n=1}^N (N\!-\!2n\!+\!1)^2(\hat{\bk}_n\!-\!\hat{\bk}_{n-1})^2\bigg]\right\rangle.
\end{equation}
Assuming that the atoms scatter isotropically, one has $\smash{\langle\hat{\bk}_n\hat{\bk}_{n-1}\rangle}=0$, which leads to
\begin{equation}
c(N)=\exp\!\bigg[-\!\frac{(k\bar{v}\tau)^2(N^3-N)}{3}\bigg].
\label{cN_coldatoms}
\end{equation}
This relation defines the CBS contrast for a given sequence of $N$ scattering events.
The full CBS contrast follows by multiplying $c(N)$ by the diffusive weight $1/N^{3/2}$  of the sequence and 
summing over all possible sequence lengths: $\sum_{N=1}^\infty c(N)/N^{3/2}$.
We finally obtain the full CBS contrast $C$ by dividing this quantity by the incoherent multiple scattering background, $\sum_{N=1}^\infty 1/N^{3/2}$:
\begin{equation}
\label{CBS_contrast_def}
C=\frac{\sum_{N=1}^\infty c(N)/N^{3/2}}{\sum_{N=1}^\infty 1/N^{3/2}},
\end{equation}
which can be conveniently rewritten as
\begin{equation}
\label{CBS_contrast_coldatoms}
C=1-\frac{\sum_{N=1}^\infty [1-\exp[-{(k\bar{v}\tau)^2(N^3-N)}/{3}]]/N^{3/2}}{\sum_{N=1}^\infty 1/N^{3/2}}.
\end{equation}
In the usual multiple scattering experiments with cold vapors, the atoms are excited in the vicinity of resonance so to make the scattering cross section large enough. In this regime, the scattering time is mainly governed by the time delay accumulated by light at each scattering event, i.e., $\tau\simeq\Gamma^{-1}$, where $\Gamma$ is the spectral width of the atomic transition. 
At the low  temperatures where these experiments operate, one typically has $k\bar{v}/\Gamma\ll1$. The sum in the numerator of Eq. (\ref{CBS_contrast_coldatoms}) can then be approximated by an integral, whose evaluation gives:
\begin{equation}
C=1-\frac{2\Gamma(5/6)}{3^{1/6}\zeta(3/2)}\left(\frac{k\bar{v}}{\Gamma}\right)^{1/3},
\label{CBS_contrast_coldatoms_fin}
\end{equation}
where $\Gamma(x)$ and $\zeta(x)$ are respectively the Euler gamma and Riemann zeta functions. Atomic thermal motion therefore leads to a reduction of the CBS interference contrast. Equation (\ref{CBS_contrast_coldatoms_fin}) also provides a simple rule of thumb to estimate the impact of thermal decoherence: in order for the CBS peak to be well contrasted, the typical time $\lambda/\bar{v}$ the atoms take to move over a wavelength $\lambda=2\pi/k$ should be much longer that the time delay $\Gamma^{-1}$ associated with a scattering process. 
As a remark, however, let us mention that while the $(k\bar{v}/\Gamma)^{1/3}$ scaling in Eq. (\ref{CBS_contrast_coldatoms_fin}) is accurate, its numerical prefactor is only qualitative due to two main approximations made in the above reasoning. The first is the assumption of a fixed, deterministic time $\tau$ between scattering events, and the second is the inaccurate description of low scattering orders by a diffusive law in Eq. (\ref{CBS_contrast_def}). These two aspects will be properly addressed in the calculation of the next section.

While Eq. (\ref{CBS_contrast_coldatoms_fin}) was derived assuming $k\bar{v}\ll\Gamma$, 
a quick look at Eq. (\ref{cN_coldatoms}) sheds light on the fate of CBS at high temperature $k\bar{v}\gg\Gamma$. Since the decoherence accumulates very rapidly with $N$ \cite{Labeyrie06}, in this regime the CBS peak becomes mainly dominated by low scattering orders. 
Even for $N=2$ though, the contrast decays very fast with $k\bar{v}\tau$. 
At first sight,  this suggests that detecting weak localization of light in a hot atomic vapor is hopeless. 
A closer look at the microscopic expression of $\tau$, however, reveals a possible strategy to limit the impact of thermal decoherence. We recall that $\tau$ is the average delay time accumulated by light during a scattering process. This delay reflects the tendency of light  to be ``trapped'' within the atom when the latter is excited near resonance (corresponding to a large $\tau$), whereas far from resonance photon scattering becomes nearly instantaneous. Note that this phenomenon is specific for resonant scattering, as involved in the atom-photon interaction, and is absent in classical setups where CBS arises, e.g., from milk particles, teflon or polystyrene beads \cite{Albada85, Wolf85}. The delay time is defined by $\tau=\partial \Phi/\partial\omega$ \cite{Wigner55}, where $\Phi$ is the phase shift acquired by the wave during the scattering process. For light scattering from two-level atoms, $\Phi$ is the phase of the $t$-matrix [see Eq. (\ref{eq:tmatrixscalar}) below], and the delay time reads \cite{Weiss18}
\begin{equation}
\tau=\frac{\Gamma/2}{\delta^2+(\Gamma/2)^2},
\end{equation}
where $\delta$ is the frequency detuning with respect to the atomic resonance. This shows that by detuning the laser  far enough from resonance, $|\delta|\gg \Gamma$, $\tau\sim \Gamma^{-1}(\Gamma/\delta)^2$ becomes much shorter than $\Gamma^{-1}$. In this regime, the dephasing factor $k\bar{v}\tau\ll 1$ and one could expect a restoration, at least partial,  of the CBS contrast. This is the idea that we explore in the next section.

\section{Hot atoms}
\label{sec_scalar_CBS}

\subsection{Model and hypotheses}

We now turn our attention to a hot atomic vapor, for which $k\bar{v}\gg\Gamma$. From now on, we make use of a microscopic, more quantitative description of coherent backscattering. Again we consider a semi-infinite
 medium illuminated  at normal incidence by a plane-wave beam of  incident wave vector now denoted by $\bk_\text{in}$, with  $|\bk_\text{in}|=k=\omega/c$. 
In general, the beam can also be polarized, which we describe by a complex, unit polarization vector $\eps$; see Fig. \ref{fig:CBS_scheme}(b). 
 The medium is assumed to be uniformly filled with an atomic cloud of density $\rho$, excited around a resonance of frequency $\omega_0$ and spectral width $\Gamma$. 
As before, we model the distribution of atom velocities by the Gaussian law (\ref{velocity_distrib}).
As mentioned in Sec. \ref{sec_cold_atoms}, at high temperature the impact of thermal motion can be reduced by operating at large detuning $\delta=\omega-\omega_0$. Therefore, in the following we assume
\begin{equation}
|\delta|\gg k\bar{v}\gg \Gamma.
\label{large_delta_cond}
\end{equation}
Note that this regime drastically differs from the one considered in experiments with cold atoms, where one typically has $|\delta|,k\bar{v}\ll \Gamma$.
To calculate the CBS signal, we introduce the bistatic coefficient $\gamma$, defined as the ratio of the light flux scattered in a direction $\bk_\text{out}$ in reflection
to the flux of the incident plane wave
 \cite{Ishimaru97},
\begin{equation}
\gamma=\frac{4\pi}{A}\left\langle\frac{d\sigma}{d\Omega}(\eps\,
\bk_\text{in}\to
\eps'
\bk_\text{out})\right\rangle.
\label{bistatic_coeff}
\end{equation}
Here $A$ is the illuminated surface and $d\sigma/d\Omega$ is the differential scattering cross section of the vapor
 in a solid angle $\Omega$ around  direction $\bk_\text{out}$. 
Brackets refer to both averaging over external degrees of freedom (atomic positions, atomic velocities)
 and internal degrees of freedom (population of atomic energy levels). 
This definition also accounts for the possibility to measure the reflection signal in a given polarization direction, parametrized by the unit vector $\eps'$.

\subsection{Coherent backscattering in the scalar approximation}
\label{CBS_calc}

As a proof of principle, we first examine coherent backscattering at high temperature  within a scalar description. The latter amounts to both treating the atoms as classical dipoles and discarding the polarization degrees of freedom. 
As pointed out in Sec. \ref{sec_cold_atoms},
at high temperature Doppler  phase shifts tend to quickly accumulate with the number of scattering events. It follows that when $k\bar{v}\gg\Gamma$, one expects CBS to be essentially controlled by double scattering contributions. Below we will confirm this conclusion by Monte Carlo numerical simulations.

We thus focus our attention on double scattering CBS, whose interference mechanism is illustrated in Fig. \ref{fig:CBS_scheme}(b).
In contrast to the previous section, for convenience here we describe the thermal motion in the frequency domain rather than in the temporal domain.
The optical beam is scattered from two atoms moving at velocities $\bv_1$ and $\bv_2$, and is eventually detected at a distant point $\boldsymbol{R}$ in the far field, parametrized by 
the angle $\Delta\theta\ll1$ away from backscattering. After each  scattering event, the optical field gets multiplied by the so-called atomic $t$-matrix. Within a scalar description, the latter reduces to a complex scalar number given, at frequency $\omega$, by
\begin{equation}
t_s(\omega)=\frac{4\pi}{k}\frac{\Gamma/2}{\delta+i\Gamma/2}.
\label{eq:tmatrixscalar}
\end{equation} 
In the presence of thermal motion, the frequency in the $t$-matrix is, in general, modified by a Doppler shift in the rest frame of the atom.
At large detuning $|\delta|\gg k\bar{v}$, however, this modification can be neglected. For instance, the amplitude of the incoming wave on atom 1 is, after scattering, multiplied by $t_s(\omega-\bk_\text{in}\cdot\bv_1)= t_s(\omega)[1+\mathcal{O}(k\bar{v}/\delta)]\simeq t_s(\omega)$ at leading order. For the same reason, at large detuning, the mean free path $\ell=4\pi/(\rho |t_s(\omega)|^2)$ remains very close to its zero-temperature value,
\begin{equation}
\frac{1}{\ell}\simeq \frac{4\pi\rho}{k^2}\frac{(\Gamma/2)^2}{\delta^2+(\Gamma/2)^2}.
\label{mfp_scalar}
\end{equation}
For the process of Fig. \ref{fig:CBS_scheme}(b), the bistatic coefficient  thus reads
\begin{align}
\label{gammaCBS}
	\gamma_\text{C} &\simeq \frac{4\pi}{A}R^2 \rho\int d\br_1d\br_2 d\bv_1 d\bv_2 p(\bv_1) p(\bv_2)|t_s(\omega)|^4 \nonumber \\
	 &\times\Psi_\text{in}(\br_1,\omega)\Psi_\text{in}^*(\br_2,\omega)
	 \Psi_\text{out}(\boldsymbol{R},\omega')\Psi_\text{out}^*(\boldsymbol{R},\omega'')\nonumber\\[0.2\baselineskip]
	 & \times\overline{G}(\br_2,\br_1,\omega_1)\overline{G}^*(\br_1,\br_2,\omega_2).
\end{align}
In this formulation, $\Psi_\text{in}(\br_1,\omega)$ refers to the field incident on atom 1,
\begin{equation}
\Psi_\text{in}(\br_1,\omega)=
\exp({i\bk_\text{in}\cdot\br_1-z_1/2\ell}),
\label{Psiin}
\end{equation}
with an analogous expression for the field $\Psi_\text{in}(\br_2,\omega)$ incident on atom 2. Notice that Eq. (\ref{Psiin}) accounts for the attenuation of the light field between the interface and the atom, governed by the mean free path. 
The probability amplitude to  propagate from atom 1 to atom 2 is given by the disorder-averaged Green's function, 
\begin{equation}
\label{G12}
\overline{G}(\br_2,\br_1,\omega_1)=
\frac{\exp[i(|\bk(\omega_1)|+i/2\ell)|\br_2-\br_1|]}
{4\pi |\br_2-\br_1|}.
\end{equation}
In the rest frame of the atom, this field oscillates at the Doppler-shifted frequency  $\omega_{1}=\omega+\bs{v}_1\cdot(\bk-\bk_\text{in})$. The reversed wave propagating from atom 2 to atom 1 is controlled by the Green's function $\smash{\overline{G}^*(\br_1,\br_2,\omega_2)}$, and oscillates at frequency $\omega_{2}=\omega+\bs{v}_2\cdot(-\bk-\bk_\text{in})$.
Finally, the (spherical) wave emitted from atom 2 and detected in reflection is given by
\begin{equation}
\Psi_\text{out}(\boldsymbol{R},\omega')=
\frac{\exp[i|\bk_\text{out}(\omega')||\boldsymbol{R}-\br_2|-z_2/2\ell]}
{4\pi |\boldsymbol{R}-\br_2|},
\label{PsiOut}
\end{equation}
where we used that $\cos\Delta\theta\simeq 1$ in writing the attenuation factor. This field oscillates at frequency $\omega'=\omega+\bs{v}_1\cdot(\bk-\bk_\text{in})+\bs{v}_2\cdot(\bk_\text{out}-\bk)$, while the field $\Psi_\text{out}(\boldsymbol{R},\omega'')$ emitted by atom 1  oscillates at frequency $\omega''=\omega+\bs{v}_1\cdot(\bk+\bk_\text{out})+\bs{v}_2\cdot(-\bk_\text{in}-\bk)$. 
In the vicinity of backscattering, $\omega'\simeq\omega''$, which is an equality that we will assume from now on.
By making use of $|t_s(\omega)|^4=(4\pi)^2/\rho^2\ell^2$ and expanding Eq. (\ref{PsiOut}) in the far field $R\gg r_1, r_2$, we obtain
\begin{align}
\label{gammaCBS_2}
	\gamma_\text{C} &\simeq \frac{1}{4\pi A\ell^2} \int d\br_1d\br_2 d\bv_1 d\bv_2 p(\bv_1) p(\bv_2) \nonumber \\
	 &\times\exp[i(\br_1\!-\!\br_2)\cdot(\bk_\text{in}(\omega)\!+\!\bk_\text{out}(\omega'))-(z_1\!+\!z_2)/\ell]
	 \nonumber\\[0.2\baselineskip]
	 & \times\frac{\exp[i|\br_1\!-\!\br_2|(k(\omega_1)\!-\!k(\omega_2))\!-\!|\br_1\!-\!\br_2|/\ell]}{|\br_1-\br_2|^2}.
\end{align}
At this stage, it is worthwhile to note that all wave vectors involved in this equation include the refractive index $n$ of the atomic vapor. For instance, one has  $k(\omega)={n(\omega)\omega}/{c}$ with
\begin{equation}
n(\omega)=1-\frac{\rho\, \Re\,  t_s(\omega)}{2k^2}=1-\frac{2\pi\rho}{k^3}\frac{\delta\, \Gamma/2}{\delta^2+(\Gamma/2)^2}.
\label{eq_index}
\end{equation}
With this in mind, we expand the Doppler-shifted wave vectors in Eq. (\ref{gammaCBS_2}) around the laser frequency:
\begin{equation}
k(\omega_1)\simeq k(\omega)+\bv_1\!\cdot\![\bk(\omega)\!-\!\bk_\text{in}(\omega)]\frac{\partial k}{\partial\omega}.
\label{expansion_k}
\end{equation}
The frequency  derivative of  $k(\omega)=n(\omega)\omega/c$ is mainly controlled by the frequency variation of the refractive index (\ref{eq_index}), which at large detuning gives
\begin{equation}
\frac{\partial k}{\partial\omega}\simeq\frac{\pi\rho\Gamma}{k^2\delta^2}\simeq\frac{1}{\Gamma\ell}.
\label{dkdomega}
\end{equation}
Note that Eqs. (\ref{expansion_k}) and (\ref{dkdomega}) provide an explicit estimate of the Doppler correction :
\begin{equation}
\frac{k(\omega_1)-k(\omega)}{k(\omega)}\sim \rho k^{-3}\frac{k\bar{v}}{\delta}\frac{\Gamma}{\delta}\ll1.
\end{equation}
This inequality a posteriori  validates the Taylor expansion (\ref{expansion_k}), and also underlines the interest of working at large detuning for limiting the impact of Doppler shifts.

Next we insert Eq. (\ref{expansion_k}) and the analogous expansions for $k(\omega_2)$, $\bk_\text{in}(\omega')$ and $\bk_\text{out}(\omega')$ into Eq. (\ref{gammaCBS_2}), and perform the Gaussian integrals over $\bv_1$ and $\bv_2$ using Eq. (\ref{velocity_distrib}). This leads to
\begin{align}
\label{gammaCBS_3}
	\gamma_\text{C} &\simeq \frac{1}{4\pi A\ell^2} \int d\br_1 d\br_2
	\frac{\exp[-(|\br_1\!-\!\br_2|+z_1\!+\!z_2)/\ell]}{|\br_1-\br_2|^2}
	 \nonumber \\
	 &\times\exp\Big[-\frac{\bar{v}^2|\bk-\bk_\text{in}|^2}{2(\Gamma\ell)^2}(z_2-z_1+|\br_1\!-\!\br_2|)^2\Big]
	 \nonumber\\[0.2\baselineskip]
	 & \times\exp\Big[-\frac{\bar{v}^2|\bk+\bk_\text{in}|^2}{2(\Gamma\ell)^2}(z_2-z_1-|\br_1\!-\!\br_2|)^2\Big]
	 \nonumber\\[0.2\baselineskip]
	 &\times\exp[i(\br_1\!-\!\br_2)\!\cdot\!(\bk_\text{in}+\bk_\text{out})],
\end{align}
where all wave vectors are now evaluated at the laser frequency $\omega$.
Introducing the polar angle $\theta$, see Fig. \ref{fig:CBS_scheme}(b), and $x=|\br_1-\br_2|/\ell$, Eq. (\ref{gammaCBS_3}) can be rewritten as
 \begin{align}
\label{gammaCBS_4}
	\gamma_\text{C} &\simeq \frac{1}{2} \int_0^\infty d x \int_0^{\pi/2} \!\!\sin\theta d\theta
	\exp[-x(1+\cos\theta)]
	 \nonumber \\
	 &\times J_0(k\ell x\sin\theta\Delta\theta)\exp[-2\left(k\bar{v}/\Gamma\right)^2x^2\sin^2\theta].
\end{align}
This is the general form of the double scattering  CBS contribution to the bistatic coefficient, valid at large detuning $|\delta|\gg k\bar{v}$. Note that this expression holds for arbitrary value of the ratio $k\bar{v}/\Gamma$, even though higher scattering order also come into play when $k\bar{v}\ll\Gamma$, as explained in Sec. \ref{sec_cold_atoms}.

In the rest of this section, we normalize for convenience the CBS contribution to the corresponding  incoherent, double scattering background signal, denoted by $\gamma_\text{B}$. Notice, however, that $\gamma_\text{C}/\gamma_\text{B}$ coincides with the CBS contrast only when other scattering orders, in particular single scattering, give negligible contributions to the background. Configurations allowing us to satisfy this condition in practice will be discussed in Secs. \ref{sec:quantum_structure} and \ref{sec:geometry}.
The double scattering background $\gamma_\text{B}$  is deduced from the process in Fig. \ref{fig:CBS_scheme}(b) by reversing one of the two paths. This amounts to simply setting $\Delta\theta=0$ and $\bar{v}=0$ in Eq. (\ref{gammaCBS_4}). Integrals over $r$ and $\theta$ can then be readily performed to give
\begin{equation}
\gamma_\text{B}=\frac{\ln 2}{2}.
\label{gammaB_scalar}
\end{equation}

\subsection{CBS maximum and angular profile}
\label{sec:CBS_constrastwidth}

\begin{figure}
\includegraphics[scale=0.85]{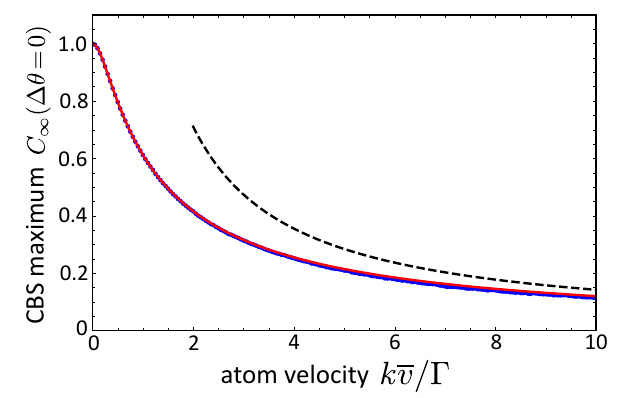}
\caption{
CBS peak maximum  normalized to the double scattering incoherent background, $C_\infty(\Delta\theta=0)=\gamma_\text{C}(\Delta\theta=0)/\gamma_\text{B}$, as a function of $k\bar{v}/\Gamma$, in the large detuning regime and in the scalar approximation. The solid red  curve is Eq. (\ref{gammaCBS_4}), and the dashed curve is the high-temperature asymptotic law $(\pi^{3/2}/2^{5/2}\ln 2) \Gamma/k\bar{v}$.
Blue dots are obtained from Monte Carlo simulations for double scattering, taking $\delta=50\Gamma$.
}
\label{fig:CBS_contrast}
\end{figure}

We first examine the CBS peak maximum, $C_\infty(\Delta\theta=0)=\gamma_\text{C}(\Delta\theta=0)/\gamma_\text{B}$, where the $\infty$ index refers to the semi-infinite geometry.
We show this quantity in Fig. \ref{fig:CBS_contrast} as a function of $k\bar{v}/\Gamma$. As expected,  $C_\infty(0)$ becomes close to unity in the low temperature limit $k\bar{v}\ll\Gamma$ : Doppler shifts vanish and the coherent and incoherent contributions become equal to each other. 
At high temperature, $C_\infty(0)$ is reduced  but we observe that this reduction is only \emph{algebraic}. This property, which is one of the main results of the paper, can be explicitly shown by evaluating Eq. (\ref{gammaCBS_4}) 
in the limit $k\bar{v}\gg \Gamma$. The integrals are then dominated by the saddle point $\theta=0$. Expanding the integrand around that point, we obtain
\begin{eqnarray}
\label{CBS_asymptotic}
C_\infty(\Delta\theta)=\frac{\gamma_\text{C}(\Delta\theta)}{\gamma_\text{B}}&\underset{k\bar{v}\gg\Gamma}{\simeq}&
\frac{\pi^{3/2}}{2^{5/2}\ln 2}\frac{\Gamma}{k\bar{v}}\,
F\Big(\frac{\Delta\theta\Gamma\ell}{4\bar{v}}\Big),
\end{eqnarray}
where $F(x)=\exp(-x^2)I_0(x^2)$ with $I_0$ is the modified Bessel function of the first kind of zero order. 
At $\Delta\theta=0$, this provides $C_\infty(0)\simeq 1.42 \Gamma/k\bar{v}$, which decays as $1/\sqrt{T}$. This asymptotic limit is displayed in Fig. \ref{fig:CBS_contrast} (dashed curve). 
Noticeably, in the regime of large detuning considered here the contrast factor $C_\infty(0)$ in turn becomes \emph{independent} of $\delta$. This saturation implies that  Doppler shifts can only be \emph{partially} counterbalanced by an increase of the detuning. This is a marked difference with the mirror-assisted CBS effect considered in \cite{Cherroret19}, where increasing $\delta$ allows one to effectively reproduce the physics of a zero-temperature atomic gas.

\begin{figure}[h]
\includegraphics[scale=0.84]{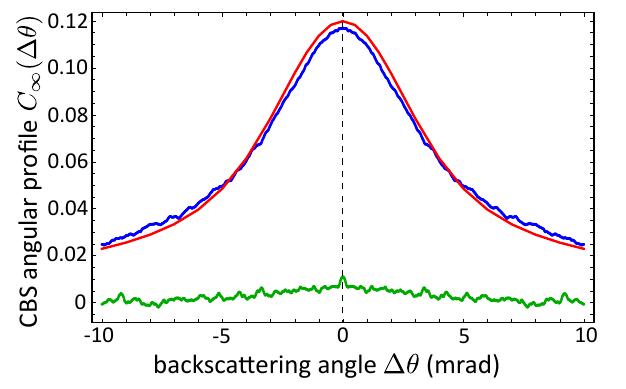}
\caption{
Angular profile of the CBS peak at fixed $k\bar{v}=10\Gamma$, $\delta= 50\Gamma$ and $k\ell=10000$. The solid red curve shows $C_\infty(\Delta\theta)$ computed using Eqs. (\ref{gammaCBS_4}) and (\ref{gammaB_scalar}), and the solid blue curve is the corresponding result obtained from Monte Carlo simulations for double scattering. The lower solid green curve  is the triple scattering CBS contribution computed from numerical simulations (precisely $\gamma_\text{C}(\Delta\theta)/\gamma_\text{B}$, with $\gamma_\text{C}$ and $\gamma_\text{B}$ now the bistatic coefficients for triple scattering).
}
\label{fig:CBS_profile}
\end{figure}
Figure \ref{fig:CBS_profile} then shows the full angular profile $C_\infty(\Delta\theta)$ of the CBS peak, at fixed  $k\bar{v}=10\Gamma$ and $\delta= 50\Gamma$. In such a high temperature  and large detuning regime, a good estimate of the CBS angular width $\Delta\theta_\text{C}$ is provided by Eq. (\ref{CBS_asymptotic}),
\begin{equation}
\Delta\theta_\text{C}\simeq \frac{4}{k\ell}\frac{k\bar{v}}{\Gamma}\simeq 4\pi\rho k^{-3}\frac{k\bar{v}}{\Gamma}\left(\frac{\Gamma}{\delta}\right)^2.
\end{equation}
Observe that  the width is controlled by the competition between two terms: 
the disorder strength $1/k\ell\ll 1$, which tends to make the width narrow, and  the Doppler factor $k\bar{v}/\Gamma\gg1$, which makes it broad.  Qualitatively, this broadening originates from the saddle point $\theta=0$ selected by the Doppler factor in Eq. (\ref{gammaCBS_4}). The latter imposes a small interatom spacing  in the plane of the interface, and therefore a broad CBS ``fringe''.
In any case, because of this competition a practical  observation of coherent backscattering in this regime may require a  careful choice of the detuning and  temperature, so that $\Delta\theta_\text{C}$ does not fall below the angular resolution of the apparatus. For the parameters chosen in Fig. \ref{fig:CBS_profile}, the CBS peak is rather broad.

\subsection{Numerical test}

In order to test the validity of our theoretical approach, we have performed Monte Carlo numerical simulations of multiple scattering by the atomic vapor. This method has been widely used in the context of multiple scattering from cold atomic clouds \cite{Labeyrie03}. 
In short, it amounts to following the propagation of a photon in the effective medium by computing sequences of the type of Eq. (\ref{gammaCBS}), choosing the position of a given scatterer in a random direction and at an exponentially distributed random distance  from the previous one.
In the present case, we apply this approach to the geometry of a semi-infinite medium, taking into account the velocity distribution (\ref{velocity_distrib}) of the atoms. The Monte Carlo method allows us to numerically compute the bistatic coefficient corresponding to individual scattering orders.
We first show the numerical prediction for the double scattering CBS contrast factor in Fig. \ref{fig:CBS_contrast} (blue dots). The results are in very good agreement with the analytical calculation (\ref{gammaCBS_4}) for all values of $k\bar{v}/\Gamma$. We also show the numerical CBS angular profile in Fig. \ref{fig:CBS_profile}. Again, the results are well captured by Eq. (\ref{gammaCBS_4}).
 
An important assumption of our analytical calculations was to neglect the contribution of higher scattering orders to CBS in the regime of high temperature and detuning. To verify this assumption, we have also computed, with our Monte Carlo simulations, the contribution of triple scattering to coherent backscattering in the hot vapor. The results are displayed in Fig. \ref{fig:CBS_profile}. We find that  the triple scattering signal  is almost invisible compared to the double scattering one. This result confirms the idea that, unlike in a cold atomic cloud, in a hot vapor a rapid accumulation of Doppler phase shifts occurs and makes higher scattering orders quickly negligible.
\newline

\section{Role of the quantum level structure and polarization}
 \label{sec:quantum_structure}
 
 In light scattering by cold atoms, an important physical ingredient affecting weak localization is the quantum level structure of the atomic scatterers. In general, the quantum selection rules of a given atomic transition give rise to an imbalance between time-reversed scattering trajectories, which eventually reduces the CBS peak contrast \cite{Labeyrie1999, Mueller01, Mueller2002, Mueller2005, Labeyrie1999}. The same phenomenon is expected to arise in a hot atomic vapor, and therefore should be accounted for. This is the task we accomplish in this section. The quantum selection rules being intrinsically sensitive to light polarization, their analysis requires one to explicitly account for the vector character of light, i.e., to go beyond the scalar description of the previous section. 
For cold atoms, the theoretical description of this problem has been presented in previous works \cite{Mueller01, Mueller2002, Mueller2005, Labeyrie03}. Below we apply this formalism to the calculation of the reflection intensity in a hot vapor, referring the reader to the aforementioned references for more details. 
We first provide the general expressions for the coherent and incoherent double scattering signals, as well as for the single-scattering contribution, which also affects the CBS contrast. We then apply these results to the concrete case of a hot alkali vapor excited near the D1 and D2 lines, and provide a reliable estimate of the contrast of the coherent backscattering peak in a polarization configuration where single scattering vanishes.

\subsection{Double scattering}
\label{sec_quantumstructure}

From now on, we assume that the light excites a two-level atomic transition involving a ground state and an excited state of total angular momenta $J$ and $J_e$, respectively. The Zeeman degeneracy of these states is controlled by magnetic quantum numbers $m$ and $m_e$, which fulfill $-J\leq m\leq J$ and $-J_e\leq m_e\leq J_e$.
Scattering from an atom with such a level structure is characterized by the $t$-matrix,
\begin{equation}
t_{ij}=t(\omega)\langle J m'|d_i d_j|Jm\rangle
\end{equation}
where $|J,m\rangle$ refer to the magnetic substates of the ground state, $\boldsymbol{d}$ is the reduced atomic dipole operator, and 
\begin{equation}
t(\omega)=\frac{6\pi}{k}\frac{\Gamma/2}{\delta+i\Gamma/2}.
\label{eq:tmatrix}
\end{equation} 
With respect to the scalar description, the disorder-averaged Green's function (\ref{G12}) is now a tensor given by $\overline{\boldsymbol{G}}(\br_2,\br_1,\omega_1)=\overline{{G}}(\br_2,\br_1,\omega_1) \boldsymbol{\Delta}$, where $\Delta_{ij}=\delta_{ij}-\hat{r}_i\hat{r}_j$ is the projector onto the plane perpendicular to the  vector $\br=\br_2-\br_1$ joining the two atoms. The scalar Green's function $\overline{{G}}(\br_2,\br_1,\omega_1) $ is still defined by Eq. (\ref{G12}), but with a mean free path now given by
\begin{equation}
\frac{1}{\ell}= M_J\frac{6\pi\rho}{k^2}\frac{(\Gamma/2)^2}{\delta^2+(\Gamma/2)^2},
\label{mfp_def_vec}
\end{equation}
with $M_J=(2J_e+1)/[3(2J+1)]$ the ratio of level multiplicities.
With these ingredients, the CBS signal detected in a polarization direction $\eps'$ and for an incident beam polarized along $\eps$ remains given by Eq. (\ref{gammaCBS}), up to the substitution
\begin{equation}
\label{substitution_rule}
|t_s(\omega)|^4\to
\langle t_\text{dir}(\omega) t_\text{rev}^*(\omega)\rangle_\text{int},
\end{equation}
where
\begin{equation}
\label{tdir}
t_\text{dir}(\omega)=\eps'^*\cdot \boldsymbol{t}_2
\cdot \boldsymbol{\Delta}\cdot 
\boldsymbol{t}_1\cdot\eps
\end{equation}
for the direct path, and
\begin{equation}
\label{trev}
t_\text{rev}(\omega)=\eps'^*\cdot \boldsymbol{t}_1
\cdot \boldsymbol{\Delta}\cdot 
\boldsymbol{t}_2\cdot\eps,
\end{equation}
 for the reversed path.
 Here, $\boldsymbol{t}_1$ and $\boldsymbol{t}_2$ are the $t$-matrices of atom 1 and atom 2, respectively, and the dot symbol denotes tensor contraction. For instance, Eq. (\ref{tdir}) should be understood as $t_\text{dir}(\omega)= \eps'^*_i 
\times(\boldsymbol{t}_2)_{ij}\times
\boldsymbol{\Delta}_{jl}\times
(\boldsymbol{t}_1)_{lk}\times
\eps_k$, where a summation over repeated indices is implied. In Eq. (\ref{substitution_rule}), $\langle\ldots\rangle_\text{int}$ refers to a statistical average over the internal Zeeman sublevels. We take them uniformly distributed, which is a reasonable assumption in a hot vapor. As pointed out in \cite{Mueller01}, due to the non-scalar part of the atomic $t$-matrix, the tensors in Eqs (\ref{tdir}) and (\ref{trev}) do not commute in general, so that $t_\text{dir}\ne t_\text{rev}$. The average in Eq. (\ref{substitution_rule}) was calculated in \cite{Mueller01}, using irreducible representations of the rotation group. The result is 
 \begin{equation}
 \langle t_\text{dir}(\omega) t_\text{rev}^*(\omega)\rangle_\text{int}=|t(\omega)|^4 M_J^2
 P_C(\eps,\eps',\hat{\br}),
 \end{equation}
 where \cite{Mueller01}
 \begin{eqnarray}
P_C&=&
(w_1^2+w_3^2)|\eps'^*\!\cdot\!\boldsymbol\Delta\!\cdot\! \eps|^2
+2w_1w_3 (\eps\!\cdot\!\Delta\!\cdot\! \eps^*)(\eps'\!\cdot\!\boldsymbol\Delta\!\cdot\! \eps'^*)
\nonumber\\
&+&(w_1+w_3)w_2
[(\eps\!\cdot\!\eps')(\eps^*\!\cdot\!\boldsymbol\Delta\!\cdot\! \eps'^*)
\!+\!(\eps^*\!\cdot\!\eps'^*)(\eps\!\cdot\!\boldsymbol\Delta\!\cdot\! \eps')]
\nonumber\\
&
+&2w_2^2|\eps'\!\cdot\! \eps|^2.
\label{PC2_exp}
\end{eqnarray}
The coefficients $w_i$ are positive weights that dependent on the specific $J\to J_e$ transition considered. 
Applying the substitution (\ref{substitution_rule}) in Eq. (\ref{gammaCBS}) together with the definition (\ref{mfp_def_vec}) of the mean free path, we finally obtain
 \begin{align}
\label{gammaCBS_polar2}
	\gamma_\text{C} &\simeq \frac{9}{8} \int_0^\infty d x \int_0^{\pi/2} \!\!\sin\theta d\theta
	\exp[-x(1+\cos\theta)]
	P_C(\theta)
	 \nonumber \\
	 &\times J_0(k\ell x\sin\theta\Delta\theta)\exp[-2\left(k\bar{v}/\Gamma\right)^2x^2\sin^2\theta].
\end{align}
Notice that as compared to the result (\ref{gammaCBS_4}) that assumed classical dipoles and scalar light, the effects of the polarization and quantum level structure are entirely contained in the additional prefactor $9/4P_C$.
The calculation of the incoherent background contribution is the same as in the scalar case,  except for the substitution 
\begin{equation}
\label{substitution_rule2}
|t_s(\omega)|^4\to
\langle t_\text{dir}(\omega) t_\text{dir}^*(\omega)\rangle_\text{int}
\!=\!|t(\omega)|^4M_J^2P_B(\eps,\eps',\hat{\br}),
\end{equation}
 where the incoherent polarization factor is given by \cite{Mueller01}
 \begin{eqnarray}
P_B&=&
(w_1^2+w_2^2)|\eps'^*\!\cdot\!\Delta\!\cdot\! \eps|^2
+2w_1w_2 |\eps'\!\cdot\!\Delta\!\cdot\! \eps|^2
\label{PL2_exp}\\
&+&(w_1+w_2)w_3
[(\eps^*\!\cdot\!\Delta\!\cdot\! \eps)
\!+\!(\eps'^*\!\cdot\!\Delta\!\cdot\!\eps')]+2w_3^2.\nonumber
\end{eqnarray}
This readily leads to
\begin{align}
\label{gammaB_polar2}
	\gamma_\text{B}\! \simeq \frac{9}{8} \int_0^\infty\!\! d x \int_0^{\pi/2} \!\!\!\sin\theta d\theta
	\exp[-x(1\!+\!\cos\theta)]
	P_B(\theta).
\end{align}
As compared to the scalar description, notice that  Eqs. (\ref{gammaCBS_polar2}) and (\ref{gammaB_polar2}) differ even in the zero-temperature limit $k\bar{v}/\Gamma=0$, since $P_C\ne P_B$ in general. The equality is only recovered for the $J=0\to J_e=1$ transition, for which $w_1=1$ and $w_2=w_3=0$ \cite{Mueller01}.

\subsection{Single scattering}

Before applying the above results to a concrete case, let us say a word about the role of light scattering by a single atom. As single-scattering processes do not have a time-reversed counterpart, they usually decrease the CBS contrast. To avoid this, a strategy consists in choosing a polarization configuration where single scattering vanishes due to the selection rules of the atomic transition. 
In the regime (\ref{large_delta_cond}) of large detuning, the single-scattering contribution is not affected by thermal motion even in a hot vapor, and is given by 
\begin{align}
\label{gamma_single}
	\gamma_\text{S} &= \frac{4\pi}{A}R^2 \rho
	\int d\br \,
	\langle|\eps'^*\!\cdot\!\boldsymbol{t}\!\cdot\!\eps|^2\rangle_\text{int}\nonumber \\
	 &\times\Psi_\text{in}(\br,\omega)\Psi_\text{in}^*(\br,\omega)
	 \Psi_\text{out}(\boldsymbol{R},\omega')\Psi_\text{out}^*(\boldsymbol{R},\omega'),
\end{align}
where the atom on which light is scattered is located at point $\br$ and is characterized by the $t$-matrix $\boldsymbol{t}$. The internal average over the polarization term is given by \cite{Mueller01}
\begin{equation}
\langle|\eps'^*\!\cdot\!\boldsymbol{t}\!\cdot\!\eps|^2\rangle_\text{int}=|t(\omega)|^2M_JP_S(\eps,\eps'),
\end{equation}
with
\begin{equation}
P_S(\eps,\eps')=
w_1|\eps'^*\!\cdot\!\eps|^2+
w_2|\eps'\!\cdot\!\eps|^2+
w_3.
\label{PS_eq}
\end{equation}
Making use of Eqs. (\ref{Psiin}) and (\ref{PsiOut}) and performing the integral over $\br$, we obtain:
\begin{equation}
\gamma_\text{S} =
\frac{3}{4}P_S(\eps,\eps').
\end{equation}
\newline
For certain atomic transitions, corresponding to specific sets $(w_1,w_2,w_3)$, $\gamma_\text{S}$ can be canceled by properly choosing a couple of incident and detection polarizations (``polarization channel''). This method can be successfully applied to the D lines of alkali atoms, as we now show.

\subsection{Application: D lines of alkali atoms}
\label{Sec:Dlines}

We now apply the above results to the D lines of an alkali vapor, frequently used in experiments on light scattering. The D1 line refers to the transition $J=1/2\to J_e=1/2$, corresponding to weights $(w_1,w_2,w_3)=(1/3,-1/3,1/3)$, and the D2 line to the transition  $J=1/2\to J_e=3/2$, corresponding to $(w_1,w_2,w_3)=(2/3,-1/6,1/6)$ \cite{Mueller01}. 
For these two transitions, inspection of Eq. (\ref{PS_eq}) reveals that the single-scattering contribution cancels out when $|\eps'^*\!\cdot\!\eps|^2=0$ and $|\eps'\!\cdot\!\eps|^2=1$, i.e. in the so-called helicity-preserving polarization channel, denoted by $h || h$. To maximize the contrast of the CBS peak in a hot vapor, it is therefore this polarization configuration that should be used. 
Note, however, that in order to achieve such a cancellation, it is required that the hyperfine level structure of the D lines plays no role, which can be achieved by operating at a detuning larger than any hyperfine splitting \cite{Mueller2005}.  
In alkali atoms,  the hyperfine structure typically involves a splitting of the ground-state level $J=1/2$ into two hyperfine levels of angular momenta $F$ and $F+1$ and separated by a frequency $\Delta$, so that single scattering will be canceled for detuning  $|\delta|\gg \Delta$. In the example of $^{39}$K, which has a conveniently small hyperfine splitting, this condition reads $|\delta|\gg 461$MHz.

Let us now examine the  CBS contrast under these conditions. In the $h || h$ channel, Eqs. (\ref{PC2_exp}) and (\ref{PL2_exp}) give $P_C(\text{D1})=(1/9)(\sin^4\theta)$, $P_{B}(\text{D1})=(2/9)\sin^2\theta$, $P_C(\text{D2})=(\sin^2\theta/144)(25\sin^2\theta-12)$, and $P_{B}(\text{D2})=(\sin^2\theta/144)(20+9\sin^2\theta)$. 
Evaluating Eqs. (\ref{gammaCBS_polar2}) and (\ref{gammaB_polar2}) in the limit $k\bar{v}/\Gamma\gg1$, we then find a CBS contrast
\begin{equation}
C_\infty^\text{D1}\simeq\frac{3\pi^{3/2}}{32\sqrt{2}}\frac{\Gamma}{k\bar{v}}\simeq 0.37\frac{\Gamma}{k\bar{v}} 
\label{C_D1_eq}
\end{equation}
and
\begin{equation}
C_\infty^\text{D2}\simeq\frac{27\pi^{3/2}}{440\sqrt{2}}\frac{\Gamma}{k\bar{v}}\simeq 0.24\frac{\Gamma}{k\bar{v}}.
\label{C_D2_eq}
\end{equation}
These results show that the D1 line is slightly more favorable in view of a detection of weak localization in a hot vapor. Note that this is the same conclusion as in a cold atomic gas, $k\bar{v}\ll\Gamma$ , for which $C_\infty^\text{D1}=5/12\simeq 0.42$ and $C_\infty^\text{D2}=53/165\simeq 0.32$ \cite{Mueller01}. 

\section{Optimizing the geometry}
\label{sec:geometry}

\subsection{Amplification in elongated cells}

In the previous sections, we have described the temperature dependence of CBS at large detuning in a hot vapor and have discussed how single scattering could be eliminated in standard atomic transitions. This led us to estimates of the CBS contrast, i.e., Eqs. (\ref{C_D1_eq}) and (\ref{C_D2_eq}) for the D lines of alkali atoms.
This result, however, is clouded by two issues. The first one concerns the contributions of higher scattering orders (triple scattering and beyond) to the background signal. Although small, these contributions, neglected in Eqs. (\ref{C_D1_eq}) and (\ref{C_D2_eq}), would in practice further reduce the CBS contrast. A second issue remains the smallness of typical values of the CBS contrast. For instance, in the recent experiment \cite{Cherroret19}, a hot rubidium vapor such that $k\bar{v}\simeq 45\Gamma$ was used, corresponding to a CBS contrast of less than $1\%$ according to Eqs. (\ref{C_D1_eq}) and (\ref{C_D2_eq}).

These difficulties, nevertheless, can both be overcome by an adequate choice of the sample geometry. A straightforward way to reduce the impact of higher scattering orders, first, consists in using a sample of thickness $\mathcal{L}<\ell$. Indeed, in this limit an incoherent scattering process involving $N$ atoms typically gives a contribution $\propto (\mathcal{L}/\ell)^N$, the largest one thus being obtained for $N=2$ once single scattering has been eliminated.
The second issue,  on the other hand, can be tackled by noticing that at high temperature $k\bar{v}/\Gamma\gg1$ the Doppler exponential factor entering the CBS contribution, given by Eq. (\ref{gammaCBS_polar2}), naturally selects out geometric configurations such that $x^2\sin^2\theta\ll 1$. The latter correspond to situations where the two atoms are either close to each other ($x\ll1$), or nearly aligned along the direction perpendicular to the plane of the interface ($\theta\ll1$). 
This is in contrast to the background contribution (\ref{gammaB_polar2}), where no Doppler factor is present  so that such a selection does not apply. 
This property suggests an interesting strategy to enhance the contrast of the CBS peak in a hot vapor: by decreasing the thickness $\mathcal{L}$ and/or the transverse width $\mathcal{R}$ of the atomic medium (both assumed infinite so far), the incoherent background processes should be more significantly reduced than the CBS signal.

\begin{figure}
\includegraphics[scale=0.735]{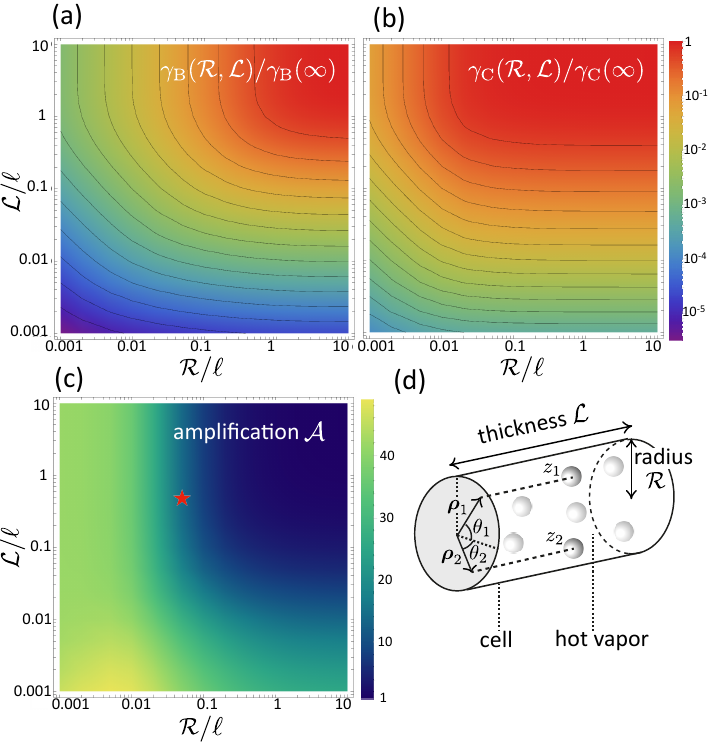}
\caption{
Calculated (a) background and (b) CBS signals in a hot vapor contained in a cylindrical cell of thickness $\mathcal{L}$ and radius ${\mathcal{R}}$, shown as density plots vs ${\mathcal{R}}/\ell$ and $\mathcal{L}/\ell$. 
Both $\gamma_\text{B}({\mathcal{R}},\mathcal{L})$ and $\gamma_\text{C}({\mathcal{R}},\mathcal{L})$ are normalized to their values in a semi-infinite medium, $\gamma_\text{B}(\infty)$ and $\gamma_\text{C}(\infty)$, respectively.
Here we set $k\bar{v}/\Gamma=40$, and the points are computed for the D1 line $J=1/2\to J_e=1/2$ in the helicity preserving channel.
Due to the Doppler dephasing factor that favors configurations with atoms close to each other and aligned with the optical axis, the CBS contribution  decreases less significantly than the background one with the cell size. (c) CBS amplification factor achieved with a cylindrical cell with respect to a semi-infinite medium [ratio of the plots (a) and (b)].  The star symbol corresponds to the parameters used in Fig \ref{fig:Final_CBS_profiles}.
(d) Cell geometry considered in this section, indicating the parametrization used in Eq. (\ref{gammaCBS_geo}).
}
\label{fig:CBS_elongatedcell}
\end{figure}

To verify this qualitative argument, we now assume that the atomic medium has the shape of a cylindrical slab of finite thickness $\mathcal{L}$ and transverse radius ${\mathcal{R}}$, as illustrated in Fig. \ref{fig:CBS_elongatedcell}(d). 
This is a realistic model for the glass cells used in experiments with hot vapors. To evaluate the bistatic coefficient in that geometry, we come back to Eq. (\ref{gammaCBS_3}), which we rewrite in terms of the projections of the atom positions $\br_1$ and $\br_2$ onto the transverse plane, $\boldsymbol{\rho}_1$ and $\boldsymbol{\rho}_2$ (of radii $\rho_1$, $\rho_2$ and polar angles $\theta_1$, $\theta_2$), and onto the $z$ axis, $z_1$ and $z_2$ [see Fig. \ref{fig:CBS_elongatedcell}(d)],
\begin{eqnarray}
\label{gammaCBS_geo}
	\gamma_\text{C}&(\mathcal{R}&,\mathcal{L},\Delta\theta) \simeq \frac{9}{16\pi A\ell^2} \int\! d\boldsymbol{\rho}_1 d\boldsymbol{\rho}_2
	dz_1 dz_2
	P_C(\boldsymbol{\rho}_1,\boldsymbol{\rho}_2,z_1,z_2)
	 \nonumber \\
	 &&
	\times	
 \frac{\exp[-(\sqrt{(\boldsymbol{\rho}_1\!-\!\boldsymbol{\rho}_2)^2\!+\!(z_1\!-\!z_2)^2}+z_1\!+\!z_2)/\ell]}{(\boldsymbol{\rho}_1\!-\!\boldsymbol{\rho}_2)^2\!+\!(z_1\!-\!z_2)^2}\nonumber\\
&&\times \exp\Big[-\frac{2\bar{v}^2k^2}{(\Gamma\ell)^2}(\boldsymbol{\rho}_1\!-\!\boldsymbol{\rho}_2)^2\Big]
{J_0(k|\boldsymbol{\rho}_1\!-\!\boldsymbol{\rho}_2|\Delta\theta)}.
\end{eqnarray}
Notice that we have included the factor $9/4P_C$ associated with the quantum level structure, as explained in Sec. \ref{sec_quantumstructure}. In this expression, the radial integrals over $\rho_1$ and $\rho_2$ run from $0$ to {$\mathcal{R}$}, the longitudinal ones over $z_1$ and $z_2$ from $0$ to {$\mathcal{L}$}, and the transverse surface of the medium is $A=\pi {\mathcal{R}^2}$. Introducing the new variables $\Delta\rho=|\boldsymbol{\rho}_1-\boldsymbol{\rho}_2|$ and $\Delta z=|z_1-z_2|$, we  simplify Eq. (\ref{gammaCBS_geo}) to
\begin{eqnarray}
\label{gammaCBS_geo_simplified}
	\gamma_\text{C}&({\mathcal{R}}&, \mathcal{L},\Delta\theta) \simeq \frac{9}{16\pi\ell} 
	\int_0^{\mathcal{L}}\!\!\! d\Delta z 
\!\int_0^{2\mathcal{R}} \!\!\!\Delta \rho\,d\Delta \rho\,J_0(k\Delta\rho \Delta\theta)
 \nonumber \\
	 &\times &
\exp\big[-\sqrt{\Delta\rho^2+\Delta z^2}/\ell\big] \exp\!\Big[\!-\frac{2\bar{v}^2k^2}{(\Gamma\ell)^2}\Delta\rho^2\!\Big] \nonumber \\
	 &\times &
 \frac{\exp(-\Delta z/\ell)-\exp[(\Delta z-2\mathcal{L})/\ell]}{\Delta\rho^2\!+\!\Delta z^2}
 P_C(\Delta\rho,\Delta z)
\nonumber\\
&\times &
\!\Big[4\arccos\Big(\frac{\Delta\rho}{2\mathcal{L}}\Big)\!-\!\frac{\Delta\rho}{\mathcal{L}}\sqrt{4\!-\!\Big(\frac{\Delta\rho}{\mathcal{L}}\Big)^2}
\Big].
\end{eqnarray}
As in the previous section, the background contribution $\gamma_\text{B}({\mathcal{R}},\mathcal{L})$ is given by the same formula up to the substitutions $\Delta\theta=0$, $\bar{v}\to 0$ and $P_C\to P_B$. 
We show, in Fig. \ref{fig:CBS_elongatedcell}, density plots of the calculated background [Fig. \ref{fig:CBS_elongatedcell}(a)] and CBS [Fig. \ref{fig:CBS_elongatedcell}(a)] bistatic coefficients normalized to their corresponding values  for a semi-infinite medium,  i.e. $\gamma_\text{C}({\mathcal{R}},\mathcal{L},0)/\gamma_\text{C}(\infty)$ 	and $\gamma_\text{B}({\mathcal{R}},\mathcal{L})/\gamma_\text{B}(\infty)$ with $\gamma(\infty)\equiv \gamma(\mathcal{R}\!=\!\infty,\mathcal{L}\!=\!\infty)$, as a function of $\mathcal{R}/\ell$ and $\mathcal{L}/\ell$. For this calculation we consider the favorable D1 line in the helicity preserving channel identified in the previous section, for which $P_C=(1/9)[\Delta\rho^2/(\Delta\rho^2+\Delta z^2)]^2$ and $P_B=(2/9)\Delta\rho^2/(\Delta\rho^2+\Delta z^2)$.
The plots confirm that in a finite sample, the  background signal is always more reduced than the CBS one when the cell size is decreased.
We also observe that CBS tends to be less affected by a decrease of the radius than by a decrease of the thickness. 
To better quantify this, we additionally show in Fig. \ref{fig:CBS_elongatedcell}(c) the ratio
\begin{equation}
\mathcal{A}=\frac{\gamma_\text{C}(\mathcal{R},\mathcal{L},0)/\gamma_\text{C}(\infty)}{\gamma_\text{B}(\mathcal{R},\mathcal{L})/\gamma_\text{B}(\infty)},
\end{equation}
 which is simply the amplification factor of the CBS contrast achieved with a cell of finite size with respect to a semi-infinite medium:
\begin{equation}
C(\Delta\theta\!=\!0)=\frac{\gamma_\text{C}(\mathcal{R},\mathcal{L},0)}{\gamma_\text{B}(\mathcal{R},\mathcal{L})}=\mathcal{A}C_\infty (\Delta\theta\!=\!0).
\end{equation}
From the figure, we find that the best amplification is obtained for a cell of small radius. If, at the same time, one wishes to maintain a large enough CBS signal, it can be preferable to operate in a configuration where $\mathcal{R}<\mathcal{L}$, i.e. with an elongated geometry.

\subsection{CBS in elongated cells: theory versus numerics}

Our final prediction, given by Eq. (\ref{gammaCBS_geo_simplified}), combined with the use of an elongated and narrow cell, $\mathcal{R}<\mathcal{L}<\ell$, constitutes the optimal configuration for observing CBS in a hot atomic vapor where $k\bar{v}/\Gamma\gg 1$. We recall that this optimal configuration is realized (i) in the regime $|\delta|\gg k\bar{v}, \Delta$, so that the Doppler effect is maximally limited and the  hyperfine structure plays no role (see Sec. \ref{Sec:Dlines}), and  (ii) in the $h\parallel h$ channel, so that single scattering is absent (note that this would no longer be true when $\delta\sim\Delta$, a regime where inelastic processes on the hyperfine structure come into play \cite{Cherroret19, Mueller2005}).

\begin{figure}
\includegraphics[scale=0.68]{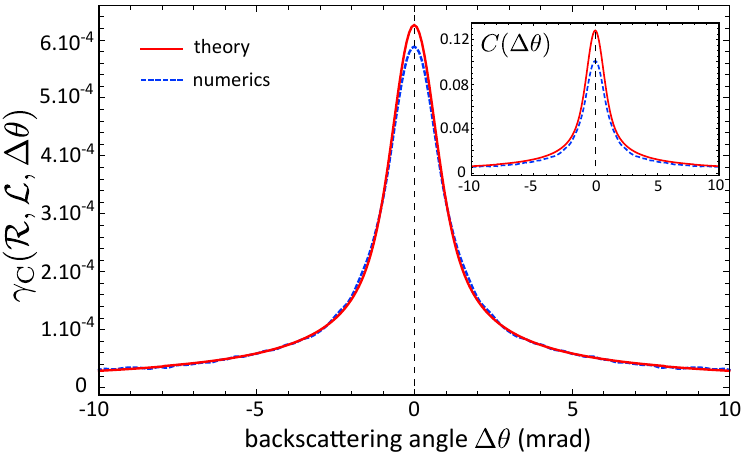}
\caption{
Calculated CBS bistatic coefficient (main panel) and contrast (inset) in a hot vapor excited near a D1 line, for an elongated cell such that $(\mathcal{R}/\ell, \mathcal{L}/\ell)=(0.05,0.5)$, corresponding to the star symbol in Fig. \ref{fig:CBS_elongatedcell}(c). 
Solid red curves are the theoretical prediction (\ref{gammaCBS_geo_simplified}), and dashed blue curves are obtained from exact Monte Carlo simulations.
Here $k\bar{v}/\Gamma=40$, $\delta=400\Gamma$ and $k\ell=160000$.}
\label{fig:Final_CBS_profiles}
\end{figure}

To conclude our analysis, we show in Fig. \ref{fig:Final_CBS_profiles} a calculation of the typical CBS bistatic coefficient, $\gamma_\text{C}(\mathcal{R},\mathcal{L},\Delta\theta)$, and in the inset the corresponding CBS contrast, $C(\Delta\theta)=\gamma_\text{C}(\mathcal{R},\mathcal{L},\Delta\theta)/\gamma_\text{B}(\mathcal{R},\mathcal{L})$ that would be effectively measured under these conditions in a hot vapor near a D1 line. 
The curves are obtained for an elongated cell such that $(\mathcal{R}/\ell, \mathcal{L}/\ell)=(0.05,0.5)$, corresponding to the star symbol in Fig. \ref{fig:CBS_elongatedcell}(c). The decoherence factor is set to $k\bar{v}/\Gamma=40$, a value close to the conditions of the experiment \cite{Cherroret19}. 
The figure shows both the theoretical prediction (\ref{gammaCBS_geo_simplified}) and the exact numerical result obtained from Monte Carlo simulations, which take into account the finiteness of the detuning
$\delta$ and multiple scattering contributions beyond double scattering. The good agreement confirms the validity of our approach. The curves also show that the CBS contrast is of several percents thanks to the geometrical confinement, to compare with the $1\%$ contrast of Eq. (\ref{C_D1_eq}) for a semi-infinite medium. 

\section{Conclusion}
\label{sec:conclusion}

In this paper, we have theoretically investigated the coherent backscattering effect for light scattered in a hot atomic vapor. This allowed us to identify the most favorable conditions for observing the mechanism of weak localization at high temperature: operating at large detuning to counteract thermal motion, working in the helicity-preserving channel to suppress single scattering, and using small and elongated atomic cells to favor the coherent backscattering signal with respect to the background one while maintaining a reasonable signal-to-noise ratio. In our approach, we have also included the quantum level structure of the atoms and identified the D1 line as a slightly better option than the D2 line. With all these conditions fulfilled, we have found that a coherent backscattering contrast of several percents could be achieved at high temperature, with the contrast being adjustable via the cell geometry. 

Although we cannot exclude the existence of other decoherence mechanisms further impacting the weak localization interference, this estimate suggests that a detection of CBS in a hot vapor is within reach of current experimental accuracy.  Let us finally mention that other strategies to enhance the CBS signal could be envisioned to counteract the impact of the quantum level structure, such as optical pumping or the use of a magnetic field to select a well-defined Zeeman transition \cite{Sigwarth04}.\newline

\begin{acknowledgments}
Part of this work was performed within the framework of the project OPTIMAL granted by the European Union by means of the Fond Européen de Développement Régional (FEDER).
\end{acknowledgments}


\begin{thebibliography}{99}

\bibitem{Albada85}
M. P. Van Albada and A. Lagendijk, 
Observation of weak localization of light in a random medium,
Phys. Rev. Lett. \textbf{55}, 2692 (1985).

\bibitem{Wolf85}
P.-E. Wolf and G. Maret, 
Weak localization and coherent backscattering of photons in disordered media,
Phys. Rev. Lett. \textbf{55}, 2696 (1985).

\bibitem{Feng91}
S. Feng and P. A. Lee, 
Mesoscopic conductors and correlations in laser speckle patterns, 
Science \textbf{251}, 633 (1991).




\bibitem{Berkovits94}
R. Berkovits and S. Feng, 
Correlations in coherent multiple scattering, 
Phys. Rep. \textbf{238}, 135 (1994).

\bibitem{Nieuwenhuizen95}
Th. M. Nieuwenhuizen and M. C. W. van Rossum,
Intensity distributions of waves transmitted through a multiple scattering medium
Phys. Rev. Lett. \textbf{74}, 2674 (1995).

\bibitem{Labeyrie1999}
G. Labeyrie, F. de Tomasi, J.-C. Bernard, C. A. Müller, C. Miniatura, and R. Kaiser,
Coherent backscattering of light by cold atoms,
Phys. Rev. Lett. \textbf{83}, 5266 (1999).



\bibitem{Keaveney12}
J. Keaveney, A. Sargsyan, U. Krohn, I. G. Hughes, D. Sarkisyan, and C. S. Adams,
Cooperative Lamb Shift in an Atomic Vapor Layer of Nanometer Thickness
Phys. Rev. Lett. \textbf{108}, 173601 (2012).

\bibitem{Javanainen14}
J. Javanainen, J. Ruostekoski, Y. Li, and S.-M. Yoo,
Shifts of a Resonance Line in a Dense Atomic Sample
Phys. Rev. Lett. \textbf{112}, 113603 (2014).

\bibitem{Jennewein16}
S. Jennewein, M. Besbes, N. J. Schilder, S. D. Jenkins, C. Sauvan, J. Ruostekoski, J.-J. Greffet, Y. R. P. Sortais, and A. Browaeys, 
Coherent Scattering of Near-Resonant Light by a Dense Microscopic Cold Atomic Cloud,
Phys. Rev. Lett. \textbf{116}, 233601 (2016).


\bibitem{Cherroret16}
N. Cherroret, D. Delande, and B. A. van Tiggelen, 
Induced dipole-dipole interactions in light diffusion from point dipoles, 
Phys. Rev. A \textbf{94}, 012702 (2016).

\bibitem{Saint-Jalm18}
R. Saint-Jalm, M. Aidelsburger, J.L. Ville, L. Corman, Z. Hadzibabic, D. Delande, S. Nascimb\`ene, N. Cherroret, J. Dalibard, and J. Beugnon, 
Resonant-light diffusion in a disordered atomic layer, 
Phys. Rev. A \textbf{97}, 061801 (2018).

\bibitem{Pellegrino14}
J. Pellegrino, R. Bourgain, S. Jennewein, Y. R. P. Sortais, A. Browaeys, S. D. Jenkins, and J. Ruostekoski, 
Observation of Suppression of Light Scattering Induced by Dipole-Dipole Interactions in a Cold-Atom Ensemble
Phys. Rev. Lett. \textbf{113}, 133602 (2014).

\bibitem{Corman17}
L. Corman, J. L. Ville, R. Saint-Jalm, M. Aidelsburger, T. Bienaimé, S. Nascimbène, J. Dalibard, and J. Beugnon,
Transmission of near-resonant light through a dense slab of cold atoms,
Phys. Rev. A \textbf{96}, 053629 (2017)


\bibitem{Guerin16}
W. Gu\'erin, M. O. Ara\'ejo, and R. Kaiser,
Subradiance in a Large Cloud of Cold Atoms,
Phys. Rev. Lett. \textbf{116}, 083601 (2016),

\bibitem{Ferioli21}
G. Ferioli, A. Glicenstein, L. Henriet, I. Ferrier-Barbut, and A. Browaeys,
Storage and Release of Subradiant Excitations in a Dense Atomic Cloud
Phys. Rev. X \textbf{11}, 021031 (2021).

\bibitem{Skipetrov14}
S. E. Skipetrov and I. M. Sokolov,
Absence of Anderson Localization of Light in a Random Ensemble of Point Scatterers
Phys. Rev. Lett. \textbf{112}, 023905 (2014).


\bibitem{Peyrot19}
T. Peyrot, Y. R. P. Sortais, J.-J. Greffet, A. Browaeys, A. Sargsyan, J. Keaveney, I. G. Hughes, and C. S. Adams,
Optical Transmission of an Atomic Vapor in the Mesoscopic Regime,
Phys. Rev. Lett. \textbf{122}, 113401 (2019).

\bibitem{Ribeiro21}
S. Ribeiro, T. F. Cutler, C. S. Adams, and S. A. Gardiner,
Collective effects in the photon statistics of thermal atomic ensembles,
Phys. Rev. A \textbf{104}, 013719 (2021).



\bibitem{Sedlacek13}
J. Sedlacek, A. Schwettmann, H. K\"ubler, and J. Shaffer,
Atom-Based Vector Microwave Electrometry Using Rubidium Rydberg Atoms in a Vapor Cell,
Phys. Rev. Lett. \textbf{111}, 063001 (2013).

\bibitem{Whiting17}
D. J. Whiting, N. \v{S}ibali\'c, J. Keaveney, C. S. Adams, and I. G. Hughes, 
Single-Photon Interference due to Motion in an Atomic Collective Excitation,
Phys. Rev. Lett. \textbf{118}, 253601 (2017).



\bibitem{Golubentsev84}
A. A. Golubentsev,
Suppression of interference effects in multiple scattering of light,
Sov. Phys. JETP \textbf{59}, 26 (1984).

\bibitem{Snieder06}
R. Snieder ,
The coherent backscattering effect for moving scatterers,
Europhys. Lett. \textbf{74}, 630 (2006).




\bibitem{Labeyrie06}
G. Labeyrie, D. Delande, R. Kaiser, and C. Miniatura,
Light transport in cold atoms and thermal decoherence,
Phys. Rev. Lett. \textbf{97}, 013004 (2006).

\bibitem{Mercadier09}
N. Mercadier, W. Guerin, M. Chevrollier, and R. Kaiser,
Lévy flights of photons in hot atomic vapours,  
Nature Physics \textbf{5}, 602 (2009).

\bibitem{Moriya16}
P. H. Moriya, R. F. Shiozaki, R. Celistrino Teixeira, C. E.
M\'aximo, N. Piovella, R. Bachelard, R. Kaiser, and Ph. W.
Courteille, Coherent backscattering of inelastic photons from
atoms and their mirror images, Phys. Rev. A \textbf{94}, 053806 (2016).

\bibitem{Piovella17}
N. Piovella, R. Celistrino Teixeira, R. Kaiser, Ph.W. Courteille, and R. Bachelard, 
Mirror-assisted coherent backscattering from the Mollow sidebands, 
Phys. Rev. A \textbf{96}, 053852 (2017).

\bibitem{Cherroret19}
N. Cherroret, M. Hemmerling, V. Nador, J.T.M. Walraven, and R. Kaiser,
Robust Coherent Transport of Light in Multilevel Hot Atomic Vapors
Phys. Rev. Lett. \textbf{122}, 183203 (2019).

\bibitem{Wigner55}
E. P. Wigner,
Lower limit for the energy derivative of the scattering phase shift, 
Phys. Rev. \textbf{98}, 145 (1955).

\bibitem{Weiss18}
P. Weiss, M. O Ara\'ujo, R. Kaiser and W. Gu\'erin,
Subradiance and radiation trapping in cold atoms,
New J. Phys. \textbf{20}, 063024 (2018).

\bibitem{Ishimaru97}
A. Ishimaru, \textit{Wave Propagation and Scattering in Random Media} (Oxford University Press, Oxford, 1997).

\bibitem{Labeyrie03}
G. Labeyrie, D. Delande, C. A. M\"uller, C. Miniatura, and R. Kaiser,
Coherent backscattering of light by an inhomogeneous cloud of cold atoms
Phys. Rev. A \textbf{67}, 033814 (2003).




\bibitem{Mueller01}
C. A. M\"uller, T. Jonckheere, C. Miniatura, and D. Delande,
Weak localization of light by cold atoms: The impact of quantum internal structure,
Phys. Rev. A \textbf{64}, 053804 (2001).

\bibitem{Mueller2002}
C. A. M\"uller and C. Miniatura,
Multiple scattering of light by atoms with internal degeneracy,
J. Phys. A: Math. Gen. \textbf{35} 10163 (2002).

\bibitem{Mueller2005}
C. A. M\"uller, C. Miniatura, D. Wilkowski, R. Kaiser, and D. Delande,
Multiple scattering of photons by atomic hyperfine multiplets,
Phys. Rev. A \textbf{72}, 053405 (2005).

\bibitem{Sigwarth04}
O. Sigwarth, G. Labeyrie, T. Jonckheere, D. Delande, R. Kaiser, and C. Miniatura,
Magnetic field enhanced coherence length in cold atomic gases,
Phys. Rev. Lett. \textbf{93}, 143906 (2004).





\end{thebibliography}
\end{document}